\documentclass[aps,twocolumn,pra,%
preprintnumbers,amsmath,amssymb,superscriptaddress]{revtex4}
\usepackage{graphicx}
\usepackage{color}
\newcommand{\bea}{\begin{eqnarray}}
\newcommand{\eea}{\end{eqnarray}}
\newcommand{\p}{\partial}
\newcommand{\Exp}[1]{\left\langle~#1~\right\rangle}

\usepackage{ulem} % 

\begin{document}

\title{Persistent Superconductor Currents in Holographic Lattices}

%\date{\today}

\author{Norihiro {\sc Iizuka}}\email[]{norihiro.iizuka@riken.jp}\email[]{iizuka@phys.sci.osaka-u.ac.jp} 
\affiliation{{\it Interdisciplinary Fundamental Physics Team, 
Interdisciplinary Theoretical Science Research Group,  RIKEN, Wako 351-0198, JAPAN }}
\affiliation{{\it Department of Physics, Osaka University, Toyonaka, Osaka 560-0043, JAPAN}}

%\author{Norihiro {\sc Iizuka}}\email[]{norihiro.iizuka@riken.jp, iizuka@phys.sci.osaka-u.ac.jp}
%\affiliation{%
%%{\it Yukawa Institute for Theoretical Physics, 
%%Kyoto University, Kyoto 606-8502, JAPAN
%{\it Interdisciplinary Fundamental Physics Team, 
%Interdisciplinary Theoretical Science Research Group,  RIKEN, Wako 351-0198, JAPAN and \\
%Department of Physics, Osaka University, Toyonaka, Osaka 560-0043, JAPAN
%}}

\author{Akihiro {\sc Ishibashi}}\email[]{akihiro@phys.kindai.ac.jp}
\affiliation{% 
{\it Department of Physics, Kinki University, Higashi-Osaka 577-8502, JAPAN
}}

\author{Kengo {\sc Maeda}}\email[]{maeda302@sic.shibaura-it.ac.jp}
\affiliation{%
{\it Faculty of Engineering,
Shibaura Institute of Technology, Saitama 330-8570, JAPAN}}

\begin{abstract}
%Beyond linear response theory, 
We consider a persistent superconductor current along the direction with no translational symmetry in a holographic gravity model. Incorporating a lattice structure into the model, we numerically construct novel solutions of hairy charged stationary black brane with momentum/rotation along the latticed direction. The lattice structure prevents the horizon from rotating, and the total momentum is only carried by matter fields outside the black brane horizon. This is consistent with the black hole rigidity theorem, and suggests that in dual field theory with lattices, superconductor currents are made up by ``composite'' fields, rather than ``fractionalized'' degrees of freedom. We also show that our solutions are consistent with the superfluid hydrodynamics. 
\end{abstract}

\maketitle

%\tableofcontents

%%%%%%%%%%%%%%%%%%%%%%%%%%%%%%%%%%%%%%%%%%%%%%%%%%%%%%

\noindent

%%%%%%%%%%%%%%%%%%%%%%%%%%%%%%%%%%%%%%%%%%%%%%%%%%
%\section{Introduction}\label{sec:intro}
%%%%%%%%%%%%%%%%%%%%%%%%%%%%%%%%%%%%%%%%%%%%%%%%%%
%{\begin{center} 
{\bf - Introduction -}
%\end{center}} 

Over the past few years a considerable number of studies have been made in applying 
the idea of holography, or the gauge/gravity duality \cite{Maldacena:1997re}, 
to strongly coupled quantum systems. 
%
%In particular, %{\color{blue} XXX the characteristic behavior of high $T_c$ superconductors or non-Fermi liquids has been  
%various 
%fascinating 
Famous examples of such are hairy black hole %new gravity 
solutions which are dual to the boundary superconductor~(see, e.g., 
\cite{Horowitz:2010gk} and references therein). %or non-Fermi liquids phenomena.   
%in the framework of linear response theory.} 
% Although a variety of conductivity or resistivity has been analyzed by applying 
% the holographic methods,  
% Although the application of such a holographic method is highly anticipated to study strongly correlated quantum systems 
% such as high $T_c$ superconductors. 
However, most of the %holographic superconductor 
models so far considered do not adequately capture essential 
features of realistic condensed matter models in the sense that 
% be far from real world condensed matter systems 
they admit translational symmetry, under which momentum is conserved. 
%and therefore ignore the effect of momentum dissipation through umklapp scattering. 
Only quite recently, %to compensate the defect, 
there have appeared some attempts to construct holographic models 
with no translational symmetry~\cite{HartnollHofman2012_17,MaedaOkamuraKoga2011, Horowitz:2012ky, Horowitz:2012gs}, 
in which infinite conductivity was confirmed in superconducting states~\cite{Iizuka:2012dk, Horowitz:2013}. 
%These analyses were, however, performed still within the linear response theory, 
%and little progress has been made at non-linear level.  
% This is because % most of the holographic models imposed translational symmetry under which momentum is conserved. 
% In order to compensate the defects, 
%
% Quite recently, to compensate the defect, several holographic models with no translational symmetry 
% have been proposed~\cite{HartnollHofman2012_17,MaedaOkamuraKoga2011, Horowitz:2012ky, Horowitz:2012gs} and 
% infinite conductivity was confirmed in superconducting states within the linear response theory~\cite{Iizuka:2012dk, Horowitz:2013}.  
%At the present stage, however, we still have a long way to understand the complete dissipation mechanism beyond 
%the linear theory.   
%  
%Beyond the linear analysis, it is highly nontrivial what kind of strongly correlated condensed matter systems 
%can be realized via a holographic gravity model. In fact, the existence of a persistent current, which is known as a 
%phenomenologically typical non-linear behavior in real world superconductors, seems to conflict with the nature of general relativity:  
% There seems to be an apparent discrepancy: 
In such holographic superconductors without translational symmetry, %On the one hand 
the persistent superconductor current is expected to have a holographic description by a stationary rotating black hole solution 
with momentum along the direction with {\it no} translational symmetry.  
% as its gravitational dual, 
On the other hand, according to black hole rigidity theorem~\cite{HE73, HIW07, MI08}, 
a stationary rotating black hole {\it must} have 
a symmetry along the direction of momentum conserved~\footnote{This aspect may be explained by the fact that 
if an asymmetric black hole solution is rotating along the direction of 
no symmetry, it loses the angular momentum by the emission of gravitational 
or electromagnetic waves~\cite{IshibashiMaeda2013}.  
Indeed, no such an asymmetric rotating black hole solution has so far been found 
except a dissipating case where entropy  
increases~\cite{Figueras:2012rb, Fischetti:2012vt}.}. % Indeed, no such a stationary rotating black hole solution has been found 
%except a dissipating case where entropy increases~\cite{Figueras:2012rb, Fischetti:2012vt}.  
%
% In other word, the momentum conservation law must be violated along the direction with no translational symmetry 
% in any stationary black hole spacetime. 
%
In this paper, we resolve the apparent conflict mentioned above 
by constructing novel black brane solutions   
{\it with momentum along the direction of no translational invariance}, which is dual to a superconducting state without dissipation. 
%in the field theory side.

In our solutions, the lattice structure prevents the horizon from rotating, and the total momentum is only carried by matter fields 
outside the black brane horizon. 
A key point is that %of our construction is that it naturally
we incorporate a lattice structure without inducing inhomogeneities and in doing so, our solutions have non-dissipating momentum along the lattice. 
To our best knowledge, this is the first example of a holographic gravity 
model that has---as a legitimate dual to a superconducting 
phase---no dissipation, and that is only made possible by taking into account 
the effects beyond the linear response theory. 
{\bf - Helical lattices from Bianchi type VII$_0$ -}%ansatz -}
%\end{center}} 

In order to introduce the lattice effects in holography, 
we will make use of the Bianchi type VII$_0$ 
geometry, characterized by the following three Killing vectors, 
\bea
&& \xi_1 = \partial_{x^2} \,, \, \, \xi_2 
= \p_{x^3} \,, \,\, \xi_3=\p_{x^1}-x^3 \p_{x^2}+x^2 \p_{x^3} \,, 
% \qquad \\ 
% && \left[ \xi_i, \xi_j \right]  = C^k_{ij} \xi_k \,,  
\eea
which form the Lie algebra, 
$\left[ \xi_i, \xi_j \right] = C^k_{ij} \xi_k $ with 
$C^1_{23}=-C^1_{32}=-1$, $C^2_{13}=-C^2_{31}=1$ and the rest $C^{i}_{jk}=0$. 
Associated with them are the following three one-forms, 
\bea
\label{invariantoneforms}
&&\omega^1=\cos(x^1)dx^2+\sin(x^1)dx^3 \,, \nonumber\\ %$ & $d\omega^1=-\omega^2\wedge\omega^3$ \\
&& \omega^2=-\sin(x^1)dx^2+\cos(x^1)dx^3 \,, \quad \omega^3=dx^1   \,,  
\quad %$ & $d\omega^2=\omega^1\wedge\omega^3$ \\ 
%\omega^3=dx^1 \,,%$ & $d\omega^3=0
\eea
each of which is invariant under all the Killing vectors
$\xi_i$.  % $\pounds_{\xi_i} \omega^j=0$.} 
% which are invariant under all of the three Killing vectors $\xi_i$.  
Using these, in this paper we make  
the following metric ansatz; 
\bea
\label{metricansatz}
ds^2 &=& -f(r) dt^2 + \frac{dr^2}{f(r)} + e^{2 v_3(r)} (\omega^3 - \Omega(r) dt)^2  \nonumber \\
&&
+  \, e^{2 v_1(r)} (\omega^1)^2  
+  e^{2 v_2(r)} (\omega^2)^2  \,.
\eea
where $f(r)$, $\Omega(r)$, $v_i(r)$, with $i=1,2,3$, are functions of 
the radial coordinate $r$ only. 
% With $\Omega(r) = 0$, this is a Bianchi type VII$_0$ of 
% the black brane solutions. 
Under more generic Bianchi type anisotropic metric ansatz, 
the near horizon geometries of static, i.e., $\Omega(r)=0$, black brane solutions,   
which admit homogeneous \footnote{Homogeneous space is the one where any two points are connected 
by the isometry of the space. %time, {\it i.e.,} Killing vectors. 
In general this isometry is not translational symmetry.} but generic anisotropic metric ansatz, 
are classified and studied in \cite{Iizuka:2012iv, Iizuka:2012pn}.
With $\Omega(r) \neq 0$, there is a flow of the geometry 
along $x^1$, {\it i.e.,} 
the black branes can ``rotate'' 
along the $x^1$ direction.  
%and there is a nonzero bulk stress-tensor component $T_{t {x^1}}$. 

Note that if $v_1(r) = v_2(r)$, then, due to 
%\bea
$(\omega^1)^2 + (\omega^2)^2 = (dx^2)^2 + (dx^3)^2$, %  \,,
%\eea
we have translational invariance along $x^1$. 
However, as long as $v_1 \neq v_2$, there is no translational invariance along $x^1$ direction, {\it i.e.,}
{$\partial_{x^1}$} 
is not a Killing vector of the geometry. Since $x^1 \to x^1 + 2 \pi n$ (with $n$ integer) is a 
discrete symmetry, there is a ``helical lattice'' structure along $x^1$ direction \cite{Donos:2012js}.  
However the metric ansatz (\ref{metricansatz}) is a homogeneous one and this homogeneity enables us to reduce the Einstein equations to a tractable set of 
ordinary differential equations.  
Note that in the homogeneous model without angular momentum, a Drude peak and nonzero resistance are found 
in a normal state \cite{Donos:2012js}, implying that momentum is generically lost due to the umklapp scattering. 
%Nevertheless, in the model, we will provide a stationary black brane solution with momentum along the helical 
%latticed direction in a superconducting state. 

Before discussing our explicit model, we can, at this stage, argue generic nature of our geometry. 
Under the metric ansatz (\ref{metricansatz}), we are concerned with 
black branes whose event horizon $H$ is given by $f(r)=0$. 
Then, the tangent vector $l$ along the null geodesics of such an $H$ is 
\bea 
l = \partial_t + \Omega_h \partial_{x^1} \,\,   \,, \qquad \Omega_h \equiv \Omega |_{r = r_h} \,, 
\eea 
where $r=r_h$ denotes the root of $f(r)=0$, {the horizon radius}. 
Then, we have, {on $H$} 
\bea
 R_{\mu\nu} l^\mu l^\nu = - 2 \Omega_h^2 (\sinh(v_1 -v_2))^2  \,.
\label{eq:Ray}
\eea   
In order to satisfy the null energy condition, {\it i.e.,} $R_{\mu\nu} l^\mu l^\nu \ge 0$, we need either $\Omega_h=0$ or $v_1 = v_2$ at the horizon~\footnote{%
Near the horizon, the Ricci scalar $R$ behaves as $R \sim \Omega_h^2 \left( \sinh \left( v_1 - v_2 \right) \right)^2/f$.
%$R \sim �_h^2 \left(sinh\left(v1 ? v2\right)\right)^2 /f$. 
So, we can reach the same conclusion from the regularity condition of the curvature.%behavior of the curvature.
%
%Regularity of the metric at the horizon can be manifestly seen by working 
%in the Eddinton-Finkelstein coordinates defined as $u:=t-\int^r f^{-1}dr$ 
%and $x^1_+:=x^1-\int^r {\Omega}f^{-1} dr$, in which $l=\p_u+\Omega_h \p_{x_+}$, 
%yielding the same result (\ref{eq:Ray}). 
%For simplicity, here and in the following we stick to the coordinates 
%introduced in~(\ref{metricansatz}). 
}.    
%$v_1 = v_2$ implies that there is a Killing vector along $\partial_{x^1}$. 
%$\Omega_h = 0$  {means the horizon} is not ``rotating''. 
This implies that it is impossible to have a rotating {horizon} along $x^1$ 
with the lattice effects.  
This is a consequence of the black hole rigidity theorem, 
which claims that under the analyticity assumption, a stationary 
rotating black hole must be 
axisymmetric~\cite{HE73,HIW07,MI08}~\footnote{This is because of the fact that for a stationary black hole, 
there must exist a null Killing vector $K$ %which is normal to
on % the event horizon
 $H$. 
% null Killing vector must exist at the horizon, 
However if the black hole is rotating, then $\partial_t$ becomes 
by definition spacelike on $H$, and therefore 
there must exist another Killing vector % $\partial_{x^1}$ 
that, combined together with $\partial_t$, provides $K$.}. 
%the horizonKilling vector 
% to combine with $\partial_t$ to yield 
% null Killing vector at the horizon. 

%%%%%%%%%%%%%%%%%%%%%%%%%%%%%%%%%%%%%%%%%%%%%%%%%%%%%%%%
%%%%%%%%%%%%%%%%%%%%%%%%%%%%%%%%%%%%%%%%%%%%%%%%%%%%%%%%
%\section{Our model}
%{\begin{center} 
{\bf - A Holographic Model -}
%\end{center}} 

The model we consider is five-dimensional action $S = \int d^5x   \sqrt{-g} \, {\cal{L}} $ where
the Lagrangian has $U(1) \times U(1)$ gauge symmetry;  
\bea
\label{action_SC}
 {\cal{L}} &=&   R+\frac{12}{L^2}-\frac{1}{4}F^2 
  -\frac{1}{4}W^2 -\, |D \Phi |^2 - m^2 |\Phi|^2  \,,  
\eea
where $R$ is Einstein-Hilbert term, $L$ represents $AdS$ scale and 
$A_\mu$, $B_\mu$ are one-form gauge potentials, and their field strengths are 
$ F=d A$, $W=dB$. 
$\Phi$ is a complex scalar field, which is charged under only the gauge potential $A_\mu$, 
but is neutral to $B_\mu$. 
Covariant derivative acting on $\Phi$ is $D_\mu=\nabla_\mu-iq A_\mu$.   

The one-form $A_\mu$ is to introduce a chemical potential.  
We make an ansatz for  {the other one-form $B$} to be proportional to type VII$_0$ Bianchi form,  
so that it induces holographic ``helical lattice'' effects. 
If we set  $B = 0$, then it makes our model the same type as \cite{Hartnoll:2008vx}.

%The Einstein equation, equations of motion for gauge bosons $A_\mu$, $B_\mu$ and a charged scalar $\Phi$ are 
%\begin{align}
%\label{Ein_SC}
%& R_{\mu\nu}=-\frac{4}{L^2}g_{\mu\nu}+\frac{1}{2}(F_{\mu\rho}{F_\nu}^\rho+W_{\mu\rho}{W_\nu}^\rho) 
%\nonumber \\
%& \qquad \quad 
%-\frac{1}{12}g_{\mu\nu}(F^2+W^2)  +\frac{m^2}{3}g_{\mu\nu}|\Phi|^2 
%\nonumber \\
%& \qquad \quad 
%+\frac{1}{2}[D_\mu \Phi(D_\nu \Phi)^\ast+D_\nu \Phi(D_\mu \Phi)^\ast] \,, \\
%\end{align}
%\begin{align}
%\label{gaugeF_SC}
%& %\nabla_\nu F^{\nu\mu}=
%\frac{1}{\sqrt{-g}}\p_\nu(\sqrt{-g} F^{\nu\mu})=iq[\Phi^\ast D^\mu\Phi-\Phi(D^\mu\Phi)^\ast] \,, \\
%\end{align}
%\begin{align}
%\label{gaugeW_SC}
%& %\nabla_\nu W^{\nu\mu}=
%\frac{1}{\sqrt{-g}}\p_\nu(\sqrt{-g} W^{\nu\mu})=0 \,, \quad  
%\end{align}
%\begin{align}
%\label{scalar_SC}
%&   
%D_\mu D^\mu \Phi=m^2\Phi \,. 
%\end{align}
We solve the field equations with the metric ansatz (\ref{metricansatz}) and 
\bea
\label{ansatzforAmu}
&&  A_\mu dx^\mu = A_{x^1}(r) \, \omega^3 + A_t(r)dt \,, \quad \\%\nonumber \\
%&=& A_1(r)d{x^1}+A_t(r)dt \,, \\
\label{ansatzforBmu}
&& B_\mu dx^\mu =  b(r) \, \omega^1 \,,   \quad \Phi = \phi(r) \,,% \quad
%&=& b(r) \left( \cos(x^1)dx^2+\sin(x^1)dx^3 \right) , \qquad 
\eea
%where $\omega^i$'s are invariant one-forms given by (\ref{invariantoneforms}), 
where we have set the phase of $\Phi$ to be zero by the gauge transformation, 
as easily checked from the equation of motion.

The equations of motion for the metric component $\xi \equiv v_1 (r) - v_2 (r)$ and $b$
%\bea
%\label{xidefinition}
%\xi \equiv v_1 (r) - v_2 (r) \,,
%\eea
are given by 
%\begin{align}
\bea
\label{eqforxi}
&& f\xi''+\{f'+f(v_1'+v_2'+v_3')\}\xi'  \nonumber \\ 
&& \quad -2 e^{-2v_3} \left(1- e^{2v_3}f^{-1}{\Omega^2}\right)\sinh 2\xi \quad \nonumber \\
&& \quad = \frac{1}{2}  {e^{-2(v_2+v_3)}}\left(1-e^{2v_3}f^{-1}{\Omega^2} \right)b^2
-\frac{1}{2}e^{-2v_1}fb'^2 , \,\,\, \quad \\
%\end{align}
%\bea
\label{eqforb}
&& fb''+\{f'+f(v_3'+v_2'-v_1') \}b' \qquad \qquad \nonumber \\
&& \qquad  \quad 
- e^{2(v_1-v_2)}\left(e^{-2v_3}-f^{-1}{\Omega^2}\right)b = 0 \,, \quad \quad 
\eea
where $'$ is the derivative with respect to $r$. 
The gauge potential $b$ plays the role of ``source term'' for $\xi$ evolution \cite{Donos:2012js}. 
It is then clear that nonzero $b$  {only introduces} the helical lattice effects, {\it i.e.,}
nonzero $\xi \neq 0$.  
There is no matter field other than $b$ which plays 
the role of source to induce the disparity between $\omega^1$ and $\omega^2$ for the metric ansatz (\ref{metricansatz}) 
in our model~\footnote{If $b=0$ and the black hole is rotating, then one can check that $\xi$ cannot have a regular solution at horizon.}.  
%under the matter field ansatz 
%for $A_\mu$ and $\Phi$, (\ref{ansatzforAmu}). 
Therefore, we seek for the solution with $b(r) \neq 0$, and therefore $\xi (r) \neq 0$; 
{\it One form $B$ with ansatz (\ref{ansatzforBmu}) is our source for 
holographic helical lattice effects. }  

%This clearly allows $b = 0$ as a trivial solution, and 
%one can therefore perform a linear perturbation analysis by first setting 
%$b=0$, obtaining a superconductor solution with $\xi=0$, and then 
%by solving the equations of motion for nonzero $b$ and $\xi$ \cite{Iizuka:2012dk}. 
%However we do not proceed in that way, as we are here interested in 
%non-linear gravitational backreaction of our superconductor current. 
% 
% Therefore we can first set $b=0$ and 
% obtain superconductor solution with $\xi=0$, and then 
% solve equations of motion for nonzero $b$ and $\xi$. This is perturbative 
% analysis for the lattice effects \cite{Iizuka:2012dk}. 
% However, instead, in this paper 
% we will solve the coupled equations fully nonlinearly for $b(r)$ and other metric components. 
% {\it i.e.,} we never regard $b(r)$ as a small perturbation, therefore, 
%Our treatment of the lattice effects is fully {\it non-perturbative}.

%{\begin{center} 
{\bf - Near horizon analysis -}
%\end{center}} 

Let us analyze the near horizon of (\ref{eqforb}) in the finite temperature case. 
% \sout{ We will set the finite temperature horizon at $r =r_h$. }
Near the horizon, the metric should have a single zero $f(r) \approx \kappa (r - r_h)~({\kappa>0})$. 
Furthermore let us assume that $v_i$ is {at least $C^2$} 
%; thus in particular, $v_i|_{r=r_h}  = \mbox{finite}$, 
%$v_i'|_{r=r_h} = \mbox{finite}$,  
%\footnote{We seek for the solution whose area-density does 
%not diverge at finite temperature, otherwise, dual field theory entropy
%density diverges at finite temperature.},
and set
\bea
\lim_{r \to r_h}\Omega(r) \equiv \Omega_h \,, \quad 
\lim_{r \to r_h}\xi(r) \equiv  \xi(r_h) \,.  
\eea
Now suppose that $\Omega_h$ is nonzero for a moment. 
Then, as the r. h. s. of (\ref{eq:Ray}) must be non-negative, 
we must impose $\xi (r_h) = 0$, and thus reduce (\ref{eqforb}) to
\begin{align}
&\kappa^2(r-r_h)^2b''+  \kappa^2 \, (r-r_h) \, b' 
+ \,  
\Omega_h^2 b \, \simeq \, 0 \,, 
\end{align}
in the $r \to r_h$ limit. 
This admits solutions 
\begin{align}
b(r) \propto (r - r_h)^{\eta_\pm} \,, \quad \eta_\pm \equiv \pm \frac{i  
\Omega_h}{\kappa}  \,. 
\end{align}
Since both solutions are {singular} at the horizon,
we have to choose their coefficients to vanish
and thus $b(r) = b'(r) = 0$ at $r \to r_h$ to obtain a smooth solution. 
However, this means 
that $b(r)$ must vanish identically in all the radius
as $b(r)$ obeys the 2nd order differential equation (\ref{eqforb}), 
and as a consequence $\xi(r) = 0$. %the lattice effects disappear.  
This itself %\sout{is OK}  
does not cause any problem  
with the rigidity theorem, since it implies that either $\Omega_h$ or $\xi$
must vanish at the horizon.
However since we are interested in constructing a holographic model
with lattices, we are interested in a solution with $b(r) \neq0$,
and this forces us to choose 
\bea
\label{rigidityforcesOmegazero}
\lim_{r \to r_h} \Omega(r) = 0 \,.   
\eea
Therefore we conclude that {\it in our set-up, the lattice effects and smoothness condition force us to have flows (or ``rotation'') only 
outside the horizon. Black brane horizon cannot be rotating.}
%{\bf Can we say this as a generic statement in more generic metric ansatz and matter contents?}
%Then there is no obstacle from the rigidity theorem
%to seeking for a solution with 
%\bea
%$\lim_{r \to r_h} b(r) \equiv b (r_h) \neq 0 \,$
%\eea
%inducing lattice effects, $\xi (r_h) \neq 0 \,$~\footnote{This effect has been investigated 
%in \cite{Donos:2012js} in a linear response theory.}.  
In addition, we choose for $A_t$ at the horizon $\lim_{r \to r_h} A_t(r) = 0 \,$, 
%\bea
%\label{nearhorizonAtzero}
%\lim_{r \to r_h} A_t(r) = 0 \,, 
%\eea
as in the static case~\footnote{%Otherwise, Euclidean Wilson loop becomes non-trivial around a vanishing circle 
%and therefore Maxwell field is singular there 
%\cite{Horowitz:2010gk}. 
This condition can be 
derived from the non-divergence of $A_\mu A^\mu$ at the horizon.}.

%Explicit equations of motion in terms of the component $f, \Omega, v_1, v_3, \xi, A_t, A_{x^1}, b, \phi$ are summarized in the Appendix A (\ref{eq_xi_SC}) - (\ref{eq_b_SC}). 

By the condition (\ref{rigidityforcesOmegazero}), 
 $\Omega\simeq \Omega_h'(r-r_h)$ near the horizon. 
Furthermore, from the explicit equations of motion in terms of the component $f, \Omega, v_1, v_3, \xi, A_t, A_{x^1}, b, \phi$, one can 
%The constraint equations at $r=r_h$,  
%combined with (\ref{nearhorizonOmega}) - (\ref{nearhorizonf}), and (\ref{nearhorizonAtzero}) 
show that there are the 9 free parameters;  
\begin{align}
\label{free_parameter_SC}
A_{x^1}(r_h), \quad A_t'(r_h), \quad \phi(r_h), \quad \xi(r_h), \quad  \quad \nonumber \\
v_1(r_h),\quad v_3(r_h), \quad \kappa, \quad \Omega_h', \quad b(r_h) \,.
\end{align}
We will tune these nonzero parameters in such a way that
% asymptotically at $r \to \infty$, 
we will have {asymptotically} AdS geometry with no
non-normalizable modes, except for $b$ field \footnote{We do have a non-normalizable mode
for $b$ and $A_\mu$ fields, but not for metric nor charged scalar field. Furthermore note that the non-normalizable mode for 
$A_\mu$ is purely constant and can be gauged away, giving zero electric field in AdS/CFT.}. 
So, we impose at $r\to \infty$, 
\bea
\lim_{r\to \infty}\xi(r)\equiv v_1(r)-v_2(r)= 0 \,.  
\label{condi:asympt:xi}
\eea
Then the lattice effect disappears and we have translational invariance restored at UV.

%By (\ref{eq_rotating_Omega_SC}), 
The asymptotic behavior of $\Omega$ is given by  
\bea 
\label{asymOmegaform}
\Omega\simeq \Omega_0+\frac{\Omega_{\rm N}}{r^4} \,.   
\eea
Then, we impose 
\bea 
\lim_{r\to \infty}\Omega(r)= 0 \,,
\quad ({\mbox{\it i.e.,} \quad \, \Omega_0=0 \,})
\label{condi:asympt:Omega}
\eea
because we are interested in a solution with no non-normalizable 
modes for $g_{\mu\nu}$.

Near the boundary $r \to \infty$, the scalar field behaves 
\begin{align}
\label{asy_scalar_SC}
 \phi \simeq C_+r^{\lambda_+}+C_-r^{\lambda_-} \,, \,\,  
 \lambda_\pm=-2\mp \sqrt{4+ { m^2L^2}   } \,. 
\end{align}
For numerics purpose, we will choose the value of mass as
$m^2L^2=-\frac{15}{ 4  }$.
In this case
$\phi\simeq C_+r^{-\frac{5}{2}}+C_-r^{-\frac{3}{2}}$, 
and both are normalizable modes. As usual, 
{we can impose the boundary condition that either $C_+$ or $C_-$ becomes zero~\cite{Hartnoll:2008vx}. 
In this paper, we shall impose} 
%\begin{align}
%\label{bc_asy_SC}
$C_-=0 \,$. 
%\end{align}
These conditions 
(\ref{condi:asympt:xi}), (\ref{condi:asympt:Omega}), and $C_-=0$ 
%(\ref{bc_asy_SC})
are achieved by tuning the parameters (\ref{free_parameter_SC}). 
Under these conditions, {following requirement   
of vanishing non-normalizable mode is satisfied,} 
\bea
\lim_{r \to \infty} \left( f(r) - \frac{r^2}{  L^2 } \right) = O\left(\frac{1}{r^2} \right) \,.
\eea

%%%%%%%%%%%%%%%%%%%%%%%%%%%%%%%%%%
%\section{Numerical solution}
%%%%%%%%%%%%%%%%%%%%%%%%%%%%%%%%%%
%{\begin{center} 
{\bf - Numerical Solution Interpolating IR and UV-}
%\end{center}} 

A numerical solution is shown in FIG. 1. This is for the parameter choice,  
%\begin{align}
%\label{parameterchoice}
%%& q = 1\,,  \, { L^2 = 2} \,, \, m^2 = -15/8 \,, \,\nonumber \\ 
%& r_h=3.033 \,, \, \kappa= 1.821\,, \,  % \nonumber \\& 
%\phi(r_h)=1  \,, \, \Omega_h' = 0.1 \,, \, \nonumber \\
%&  b(r_h) = 7.354 \,,\, A_t'(r_h) = 2.589\,,  \, A_{x^1}(r_h) = -0.7470  \,, \, \nonumber \\
%&  \xi(r_h) = -0.125 \,, \, v_1(r_h) = 0.8457 \,, \, v_3(r_h) = 1 \,,
%\end{align}
\begin{align}
\label{parameterchoice}
%& q = 1\,,  \, { L^2 = 2} \,, \, m^2 = -15/8 \,, \,\nonumber \\ 
& r_h=3.033 \,, \, \kappa= 1.821\,, \,  % \nonumber \\& 
\phi(r_h)=1  \,, \, \Omega_h' = 0.1 \,, \, \nonumber \\
&  b(r_h) = 7.354 \,,\, A_t'(r_h) = 2.589\,,  \, A_{x^1}(r_h) = -0.7470  \,, \, \nonumber \\
&  \xi(r_h) = -0.125 \,, \, v_1(r_h) = 0.8457 \,, \, v_3(r_h) = 1 \,,
\end{align}
%\begin{align}
%\label{parameterchoice}
%& q = 1\,,  \, \kappa= 1.821\,, \, { L^2 = 2} \,, \, m^2 = -15/8 \,, \, r_h=3.033 \,, \nonumber \\ 
%& \phi(r_h)=1  \,, \, \Omega_h' = 0.1 \,, \,  b(r_h) = 7.354 \,,\, A_t'(r_h) = 2.589\,,  \nonumber \\ 
%& A_{x^1}(r_h) = -0.7470  \,, \,  \xi(r_h) = -0.125 \,, \nonumber \\
%& v_1(r_h) = 0.8457 \,, \, v_3(r_h) = 1 \,.
%\end{align}
%
%\begin{figure}[ht]
%\centering
%  \includegraphics[width=8.4truecm,clip]{Fig1.eps} 
%\caption{$e^{v_1(r)}$ (solid, blue), $e^{v_2(r)}$ (dotted, red), and $e^{v_3(r)}$ (dashed, green) 
%for the parameter choice (\ref{parameterchoice}). $r = 1.635$ is a horizon.}
%\end{figure}
and 
\bea
\label{fixedpara}
q = 1\,,  \, { L^2 = 2} \,, \, m^2 = -15/8 \,.  
\eea
\begin{figure}%[ht]
\centering
  \includegraphics[width=\linewidth]{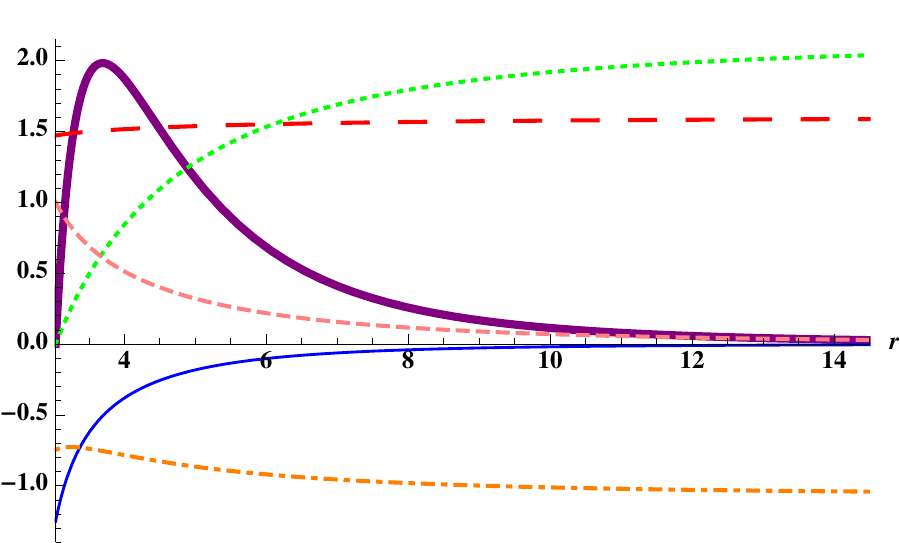} 
\caption{$100 \Omega(r)$ (thick solid, purple), {$10 \xi(r)$} (thin solid, blue), $0.5 A_t(r)$ (dotted, green), $A_{x^1}(r)$ (dotdashed, orange), $\phi(r)$ (small dashed, pink), and $0.2 b(r)$ (large dashed, red) for the parameter choice (\ref{parameterchoice}). A horizon is at $r_h=3.033$. For $\phi(r)$,  
the condition, $C_-=0$ in Eq.~(\ref{asy_scalar_SC}) is satisfied.}
\end{figure}
%\begin{figure}[ht]
%\centering
%  \includegraphics[width=8.4truecm,clip]{Fig3.eps} 
%\caption{$A_t(r)$ (solid, blue) and $A_{x^1}(r)$ (dotted, red) and $\phi(r)$ (dashed, green) 
%for the parameter choice (\ref{parameterchoice}).  
%For $\phi(r)$,  
%the condition, $C_-=0$ in Eq.~(\ref{asy_scalar_SC}) is satisfied. } 
%\end{figure}
%
In our solution, {asymptotically at $r \to \infty$, $e^{v_1} = e^{v_2} = e^{v_3} = r$} and 
the conditions~(\ref{condi:asympt:xi}), (\ref{condi:asympt:Omega})
are satisfied so that lattice disappears at the boundary and geometry 
approaches AdS metric. 
%{\color{blue} XXX removing footnote ~\footnote{The difference between $v_1(=v_2)$ and $v_3$ can be eliminated by the coordinate re-definition since $\Omega(r) \to 0$. Therefore, metric is exactly AdS$_5$ form asymptotically.}}  

The asymptotic behavior of $A_{x^1}$ is   
%\bea 
$A_{x^1}\simeq a_{x 0}+\frac{a_{x{\rm N}}}{r^2}$.  
%\eea
According to the AdS/CFT dictionary, $\sqrt{2} a_{x{\rm N}}$ corresponds to the current in the dual field theory, while 
$-2 \sqrt{2} \Omega_{\rm N}$ in (\ref{asymOmegaform})
corresponds to $\Exp{T_{t{x^1}}}$ component of the expectation value of the energy 
momentum tensor $\Exp{T_{\mu\nu}}$ on the boundary theory~\footnote{Note that we choose the convention that $16 \pi G = 1$ and $L = \sqrt{2}$, and we use the stress tensor given in \cite{Balasubramanian:1999re}.}. 
We numerically find that 
$a_{x{\rm N}}\simeq 5.873$ and $\Omega_{\rm N}\simeq 12.61$ under the boundary condition $\Omega_0=0$.  
Similarly, the asymptotic behavior of $b$ is   
%\bea 
$b\simeq b(\infty)+\frac{b_{{\rm N}}}{r^2}$, and we numerically find $b(\infty) = 8.012$.

Since $\Omega(r) \to 0$ at the horizon,
black branes are not ``rotating.''
However, as $\Omega(r) \neq 0$ between the horizon and
infinity, our solution is not static. This may be viewed that 
the matter field outside the horizon is rotating.
Note also that there is no ergosphere with respect to $\partial_t$
outside of the horizon, as $g_{tt}(r) < 0$ for all the range between
horizon to infinity. 

%%%%%%%%%%%%%%%%%%%%%%%%%%%%%%%%%%
%\section{Numerical solution}
%%%%%%%%%%%%%%%%%%%%%%%%%%%%%%%%%%
%{\begin{center} 
{\bf - Superfluid Hydrodynamics -}
%\end{center}}

Given above solution, it would be interesting to change various input parameters and 
see if there are relationship between various solutions. In TABLE I, we show some of our solutions, including 
the one in FIG 1.,  
where we varies the parameters such as; $\phi(r_h)$ (charged scalar field at the horizon), $T = \kappa/4\pi$ (temperature), 
$\mu \equiv A_t(\infty)$ (chemical potential), $b(\infty)$ and $\zeta \equiv A_{x^1}(\infty)/A_t(\infty)$ (superfluid fraction)  from the previous choice (\ref{parameterchoice}), and obtained the boundary expectation values of $\Exp{T_{t{x^1}}}$ 
and $\Exp{j_{x^1}}$ from the normalizable modes of $g_{t{x^1}}$ and $A_{x^1}$ at the boundary. 
In FIG. 2, we draw the 3D plot of $(T/\mu, b(\infty)/\mu, -\zeta)$ values of our solutions, by connecting 
vertex points which we obtained from the solutions. %fixing $\phi(r_h)$ value. 
We keep the parameters in (\ref{fixedpara}) the same. 

{\color{blue}
\begin{table}
  \centering
%\begin{tabular}{|l|l|l|l||l|l|}\hline
\begin{tabular}{|l|l|l|l|l||l|l|}\hline
$\phi(r_h)$ & $\,\,\,\,\,\,\,T$ & $\,\,\,\,\,\mu$    & $b(\infty)$  &  $\,\,- \, \zeta$ & $\,\Exp{T_{t{x^1}}}$   &  $\Exp{j_{x^1}}$  \\ \hline  
%%\,\,\,\,\,1 & 0.08138 & 4.325 & 5.927 & 0.5489 &\,\,-21.78 & \,10.07 \\ \hline 
\,\,\,\,\,1 & 0.08138 & 4.325 & 5.927 & 0.5489 &\,\,-61.60 & \,14.24 \\ \hline 
%\,\,\,\,\,1 & \,0.1048 & 4.301 & 7.181 & \,\,-20.26 & \,9.420 \\ \hline
%\,\,\,\,\,1 & \,0.1093 &  4.372 & 6.523 & \,\,-21.35 & \,9.766 \\ \hline
%%\,\,\,\,\,1 & \,0.1450 & 4.295 & 8.012 & 0.2491 & \,\,-12.61 & \,5.873 \\ \hline  
\,\,\,\,\,1 & \,0.1450 & 4.295 & 8.012 & 0.2491 & \,\,-35.67 & \,8.306 \\ \hline  
%\,\,\,\,\,1 & \,0.1492 & 4.367 & 7.249 & \,\,-13.16 & \,6.028 \\ \hline
%\,\,1/2 & 0.09389 & 3.862 & 6.866 & \,\,-4.906 & \,2.541 \\ \hline 
%%\,\,2/3 & 0.03570 & 4.071 & 4.955 & 0.7103 & -16.993 & 4.174 \\ \hline 
\,\,2/3 & 0.03570 & 4.071 & 4.955 & 0.7103 & \,\,-24.03 & 5.903 \\ \hline 
%\,\,2/3 & 0.07901 & 3.996 & 6.581 & \,\,-8.938 & \,4.473 \\ \hline
%%\,\,2/3 & \,0.1059 & 3.919 & 7.057 & 0.5018 & \,\,-8.154 & \,4.161 \\ \hline 
\,\,2/3 & \,0.1059 & 3.919 & 7.057 & 0.5018 & \,\,-23.06 & \,5.885 \\ \hline 
%\,\,3/4 & \,0.1280 & 4.007 & 6.683 & 0.2524  & \,\,-9.643 & \,4.814 \\ \hline
%%\,\,4/5 & \,0.1513 & 4.003  &  7.048 & 0.2524 & \,\,-7.237 & \,3.616 \\ \hline
\,\,4/5 & \,0.1513 & 4.003  &  7.048 & 0.2524 & \,\,-20.47 & \,5.114 \\ \hline
 \end{tabular}
  \caption{Boundary stress tensor $\Exp{T_{t{x^1}}}$ and current $\Exp{j_{x^1}}$ for various choices of output parameters;  
  charged scalar field at horizon $\phi(r_h)$, temperature $T = \kappa/4 \pi$, chemical potential $\mu = A_t(\infty)$,   
   non-normalizable source field $b(\infty)$, and superfluid fraction $\zeta = A_{x^1}(\infty)/A_t(\infty)$.} %\lim_{r \to \infty}A_{x^1}(r)/A_t(r)   $.}
  \label{tab:solutionswithvariousparameters}
\end{table}
}
%{\color{red}
%\begin{align}
%\begin{array}{|l|l|l|l|l|l|l|}\hline
% \phi(r_h) & T=\frac{\kappa}{4\pi} & \mu    &  b & -\zeta &  T_{tx}    &  j_x \\ \hline  
% 1 & 0.08138 & 4.325 & 5.927 & 0.5489 & -43.554 & 10.07 \\ \hline 
% 1 & 0.1450 & 4.295 & 8.012 & 0.2491 & -25.225 & 5.873 \\ \hline  
%2/3 & 0.03570 & 4.071 & 4.955 & 0.7103 & -16.993 & 4.174 \\ \hline 
%2/3 & 0.1059 & 3.919 & 7.057 & 0.5018 & -16.308 & 4.161 \\ \hline 
%4/5  & 0.1513 & 4.003  &  7.048 & 0.2524 & -14.474 & 3.616 \\ \hline
%\end{array}
%\end{align}
%}
%{\bf Multiply $\sqrt{2}$ for $<T_{tx^1}>$ and $<j_{x^1}>$}.
\begin{figure}%[ht]
\centering
  \includegraphics[width=7.7cm]{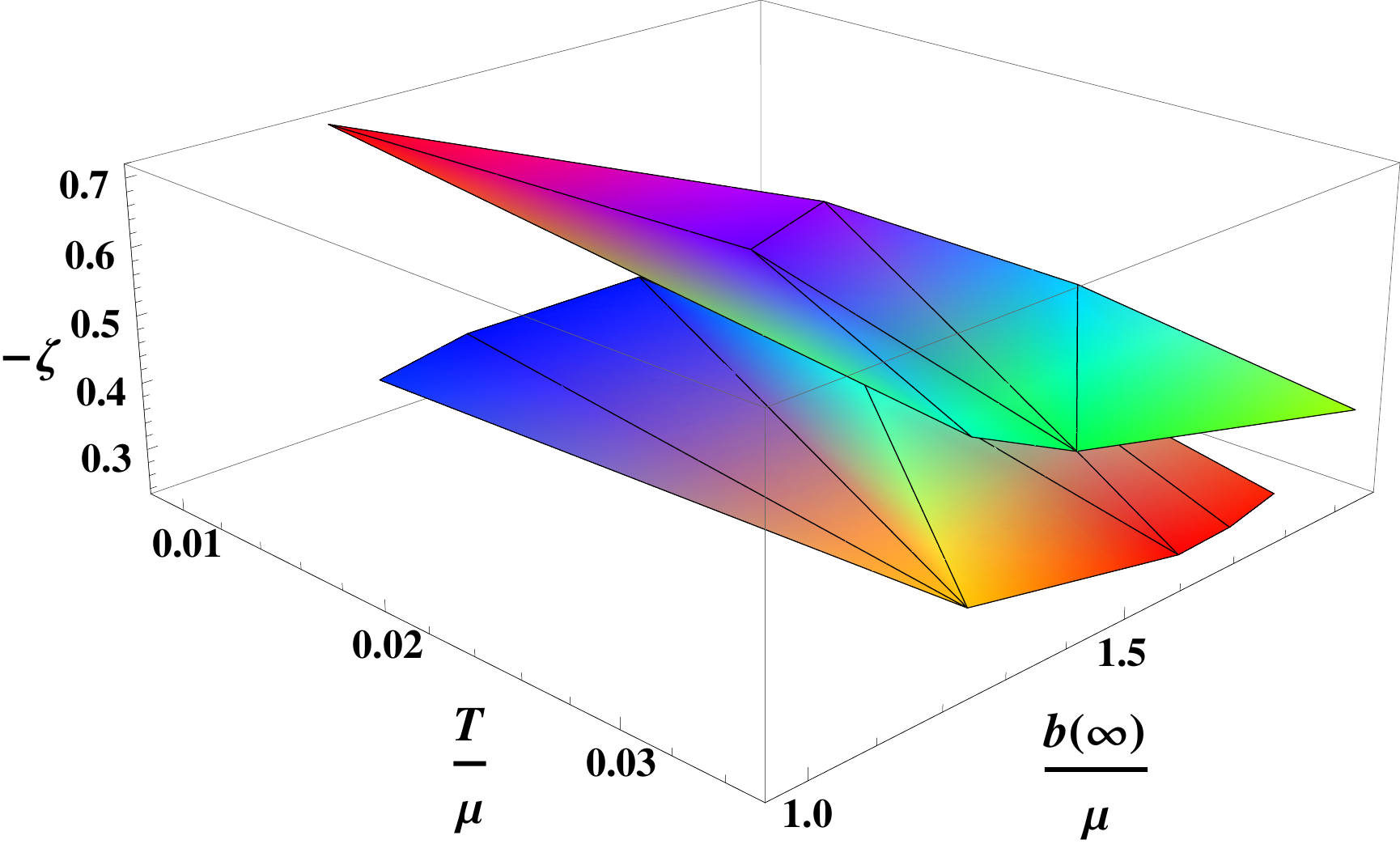} 
\caption{3D plot of dimensionless parameter $(T/\mu,$ $b(\infty)/\mu,$ $-\zeta)$ relation, fixing $\phi(r_h)$ values. 
Vertex points are obtained from numerical data. The upper surface is for fixing $\phi(r_h) = 2/3$, and 
the lower surface is for fixing $\phi(r_h) = 1$. One can clearly see the tendency that the surface exhibits a 
slope falling to the right. This implies that 
as we increase either $T/\mu$, or $b(\infty)/\mu$, or $|\zeta|$, 
the condensate VEV $\phi(r_h)$ decreases.}
\end{figure}

One can show that in all of the solutions, including the ones in the TABLE I, the relation, 
\bea
\label{hydroprediction}
\frac{\Exp{T_{t{x^1}}}}{ \mu  \Exp{j_{x^1}}} = - 1.000  \pm O(10^{-4}) \,,
\eea
holds in a very high precision, independent on the various parameter choices for $\phi(r_h)$, $T$, $\mu$, $b(\infty)$.  
%\footnote{Since one can change this value freely by radial coordinate redefinition, the value $-0.5$ itself is not much meaningful. Rather, the fact that it is a constant, independent on other parameters, is important.}. 
In the bulk viewpoint, this is a very nontrivial result obtained 
by solving Einstein equations. 
However this relation can be understood in the boundary viewpoint, %perfect 
from the superfluid hydrodynamics formula by Landau and Tisza \cite{Landau:1941, Tisza:1947zz} 
\footnote{See also \cite{Herzog:2008he,Sonner:2010yx} for related works on holographic superfluid hydrodynamics.}.    
Stress tensor and current including both normal and superfluid component are 
\bea
\label{superfluidstress}
T_{\mu\nu} &=& (\epsilon + P) u_\mu u_\nu + P \eta_{\mu\nu} + \mu \rho_s v_\mu v_\nu  \,, \quad \\
\label{superfluidcurrent}
 j_\mu &=& \rho_n u_\mu + \rho_s v_\mu  \,, 
%T^{\mu\nu} = f^2 \partial^\mu \theta \partial^\nu \theta \,, \quad  j^\mu = f^2 \partial^\mu \theta \,. 
\eea
%in the derivative expansion, 
where 
$\rho_n$, $\rho_s$, and $u_\mu$, $v_\mu$ are normal/superfluid density and velocity, 
satisfying the Josephson equation $v_\mu u^\mu = -1$. 
%$\theta$ is the phase of the superfluid/superconductor.
%Using the Josephson equation relating chemical potential and $u^\mu \partial_\mu \theta = - \mu$,   
In the absence of normal fluid velocity $u_x = 0$, (\ref{superfluidstress}) and (\ref{superfluidcurrent}) implies that 
%\bea
${T_{t{x^1}}}/{ \mu \, j_{x^1}} = v_t = - (u^t)^{-1} = -1$, which is 
%\eea
consistent with (\ref{hydroprediction}). 

Furthermore, one can see the following tendency from FIG. 2; 
the superconductivity condensate $\phi(r_h)$ %$\Exp{j_{x^1}}$ 
becomes larger,   
(1) as we lower the temperature $T/\mu$, 
(2) as we lower the lattice effects $b(\infty)/\mu$,  
(3) as we have a smaller  fraction $|\zeta|$~%condensate $\phi(r_h)$ 
\footnote{Precisely speaking, a scalar 
condensate should be read from the normalizable mode of $\phi(r)$ at the boundary, instead of 
horizon value. 
However we have seen that the latter becomes larger as the former becomes larger in all of our 
numerical solutions. Therefore 
we use $\phi(r_h)$ as ``condensate parameter''.}. 
These are expected results since high temperature and lattice effects and larger superfluid fraction 
tend to destroy superconductivity.
%, and a larger condensate corresponds to a larger superfluid fraction. 

%%%%%%%%%%%%%%%%%%%%%%%%%%%%%%%%%%%%%%%%%%%%%%%%%%
%\section{%Conclusion 
%Summary and Discussion}
%%%%%%%%%%%%%%%%%%%%%%%%%%%%%%%%%%%%%%%%%%%%%%%%%%
%
%{\begin{center} 
{\bf - Summary and Discussion -}
%\end{center}} 

In this paper, we have numerically constructed {stationary,
non-static} hairy black brane solutions, which have momentum/rotation 
along the direction of no translational invariance 
due to lattices without dissipation~\footnote{Our solutions are stationary non-rotating but not static.
This property and our metric ansatz appear to be quite similar to
the solution found in \cite{DonosGauntlett2012}, besides the obvious 
difference of matter fields considered. 
However, it should be emphasised that the solution of 
\cite{DonosGauntlett2012} contains 
a cross-term of $dt$ and $\omega^1$ or $\omega^2$
and therefore its non-staticity is due to the matter flow in a 
direction of translational symmetry $\xi_1= \partial_{x^2}$ and $\xi_2 = \partial_{x^3}$. 
In contrast, our solution, having the cross-term of $dt$ and 
$\omega^3$, contains the matter fields that flow in a direction
of {\it no} translational symmetry, $\partial_{x^1}$, and thus is a novel black brane
solution.}. 
This is dual to the persistent %non-linear 
superconductor current along the lattice direction, 
which is more realistic than previous holographic superconductor solutions.  
%
%which has no translational invariance 
%due to lattices. 
%The important point is that 
In the bulk, we consider the backreaction of $A_{x^1}$ to the metric $g_{t {x^1}}$. 
This is because the DC conductivity diverges {and} 
therefore current can be large without {external}   
electric field.  
Note that 
in our solutions, there are no source term corresponding to the external electric fields.  
%This implies that even normalizable mode $A_{x^1}$ can become large in the bulk and induce nonzero $g_{t {x^1}}$.     
%We solved this backreaction by solving Einstein equations, which is 
%%by the effects of {\it non-linear}, superconductor current, {\it i.e.,} 
%beyond the linear response. 

One of the key features in constructing our solutions is the rigidity theorem, which 
forces us one of the below choices; 
%\begin{enumerate}
%\item 
1. Holographic lattice effects survive near the horizon, but 
black brane cannot be rotating. Rotation is carried by the matter field 
surrounding black brane. 
%\item 
2. Black brane is rotating, but holographic lattice 
effects disappear.  
%\end{enumerate}
Note that in our metric ansatz, we have shown that in case of 2, 
lattice effects actually disappear {\it in all of the radius} due to the 
{regularity} 
% analyticity 
assumption, and therefore, we are forced to choose 1.  
Actually we expect that this is a generic nature. We conjecture that 
{on the gravitational action without any artificial sources,  
holographic lattices prevent black brane from rotating and momentum must
be carried by the matter fields outside.} 
%Note that this is the important conclusion we obtained by taking into account the backreaction to the metric field $g_{tx^1}$. 
%%
%{one can understand such things if and only if we take into account the backreaction of  
%$A_{x^1}$ to the metric field $g_{tx^1}$ in the bulk, which is the {\it non-linear}, superconductor current effects in the boundary.}

%{Studying this conjecture may also give some insights into 
%possible final states of a rotating black brane from the viewpoint of 
%total entropy of black brane solutions in our system (\ref{action_SC}). 
%As is well-known, a black hole carries an huge amount of entropy
%in the form of the horizon area, or the Bekenstein-Hawking entropy,
%which is typically overwhelmingly larger than the part of entropy carried 
%by matter fields outside the horizon. 
%As can be seen by inspecting existence solutions
%for rotating black holes, the horizon area of a non-rotating black hole
%is larger than that of a rotating black hole with the same mass.
%This implies that a solution with a non-rotating horizon plus rotating
%matter fields would be entropically favorable than solutions with a
%rotating horizon in the same system. To clarify this viewpoint, however, 
%we need to study, in detail, the thermodynamic properties of our solution.} 

It is very interesting to ask what it implies in dual field theory that  
holographic lattices kill the rotation of black branes. 
This might suggest that in dual field theory, 
in the presence of the lattices, ``fractionalized''  
\cite{Senthil:2003,Sachdev:2010um}  
degrees of freedom  
cannot have current along the lattice direction in the stationary limit.  
``Fractionalized'' degrees of freedom are thought to be responsible 
for the non-Fermi liquid behavior, and actually they are degrees of freedom violating 
Luttinger theorem \cite{holographicluttingerlist}. 
%\cite{Hartnoll:2010xj, Sachdev:2011ze, Huijse:2011ef, Huijse:2011hp, 
%Hartnoll:2011fn, Hartnoll:2011pp, Iqbal:2011bf, Hashimoto:2012ti}. 
It is interesting to ask {what is the implication of the rigidity theorem 
%in dual field theory side, 
and how much it gives restriction on the dynamics %current 
of ``fractionalized'' 
degrees of freedom, in condensed matter physics.}

{\bf Acknowledgments} 
The work of NI was supported by RIKEN iTHES Project. 
{This work was also supported in part by
JSPS KAKENHI Grant Number 25800143 (NI), 22540299 (AI), 23740200 (KM).}

\end{document}